\documentclass{elsart}
\usepackage{graphics,natbib,amssymb}
\begin{document}
\begin{frontmatter}

\title{Scalar Field Dark Matter, Cross Section and Planck-Scale Physics}
\author[cinves]{Tonatiuh Matos},
\ead{tmatos@fis.cinvestav.mx}
\author[cinves]{L. Arturo Ure\~{n}a-L\'{o}pez\corauthref{cor}}
\corauth[cor]{Corresponding author.}
\ead{lurena@fis.cinvestav.mx}

\address[cinves]{Departamento de F\'{\i}sica, Centro de Investigaci\'on y de Estudios Avanzados del IPN, AP 14-740, 07000 M\'exico D.F., MEXICO.}

\begin{abstract}
In recent papers we have proposed that the dark matter of the Universe could be from scalar field origin. In this letter, we find that if the scale of renormalization of the model is of order of the Planck Mass, then a scalar field $\Phi $ endowed with the scalar potential $V=V_{o}\left[ \cosh {\left( \lambda \sqrt{\kappa _{0}}\Phi \right) }-1\right] $ can be a reliable model for dark matter in galaxies. The predicted scattering cross section fits the value required for self-interacting dark matter and the additional degree of freedom of the theory is of order of hundreds of TeV.
\end{abstract}
\begin{keyword}
Cosmology, Dark Matter
\PACS 95.35.+d, 98.62.Gq, 98.80.Cq
\end{keyword}
\end{frontmatter} 

\section{Introduction}

Recent cosmological observations suggest a flat Universe in a current accelerated expansion but full with 95\% of unknown matter. 25\% of the matter in the Universe, widely known as dark matter, is responsible of the formation of the large scale structure we observe in the sky: galaxies, clusters of galaxies, voids, walls, etc. This model of dark matter has been very successful\cite{triangle}.

However, at galactic scale, numerical simulations show some discrepancies between dark-matter predictions and observations\cite{stein}. Dark matter simulations show cuspy halos of galaxies with an excess of small scale structure, while observations suggest a constant halo core density\cite{avila} and a small number of subgalactic objects\cite{avila2}. In order to solve these discrepancies, there are some proposals of scalar fields as dark matter in the literature nowadays\cite{peebles,jeremy,hu,bento,tkachev}. All of them have assumed a polynomial scalar potential up to fourth order. At the cosmological scale, the main results are cold dark matter-like behavior, growing linear perturbations, the existence of a Jeans scale and suppression of the mass power at small scales, but they depend upon initial conditions. At the galactic scale, where the self-interaction becomes important, the scattering cross section is proportional to the quartic coupling square $g^2$, which is a free parameter of the models. Even if there is a model-independent result in which a particle with a value for the scattering cross section given by (\ref{sigma}) has, straightforwardly, a mean-free path of order of 1 Mpc (see eq. (4) in\cite{bento}), the free parameters of the potential are very related, as we can see in eq. (6) of ref.\cite{bento}. Depending of the coupling $g$ the mass of the scalar field could be of order of MeV, inclusive. But nobody has determined the value of this coupling.

In this paper, we continue investigating the hypothesis of a scalar field as dark matter in the Universe with a $\cosh$ scalar potential\cite{luis,luis2,sahni}

\begin{equation}
V(\Phi ) =V_0\left[ \cosh {(\lambda \,\sqrt{\kappa _{o}} \Phi )}-1\right]
\label{cosh}
\end{equation}

\noindent with $\kappa _{0}^{-1/2}=(8\pi G)^{-1/2}=M_{Pl}/\sqrt{8\pi }=2.4\times 10^{18}\,GeV$ being the inverse square of the reduced Planck mass. Using results given in renormalization theory, we calculate the $2\rightarrow 2$ cross section at the lowest order for the potential (\ref{cosh}). It is found that the result corresponds to an effective $\phi^4$-theory in which the effective coupling $g$ is exponentially enhanced by the scale of renormalization $\Lambda$, which remains as a free parameter of the model. This result is compared with the value given for the cross section in self-interacting dark matter models. We found that if the scale of renormalization of potential (\ref{cosh}) is of order of the Planck Mass, $\Lambda ={\mathcal {O}}(M_{Pl})$, the predicted scattering cross section by mass of the scalar particles $\sigma_{2\rightarrow 2}/m_{\Phi }$ can fit the value predicted in numerical simulations of self-interacting dark matter in order to avoid high-density dark matter halos. 

\section{Scalar field dark matter with a cosh potential}

This model neither depends on initial conditions nor has problems with nucleosynthesis (provided that $\lambda >5$) and scales as radiation at early times, because of its exponential behavior. It scales as cold dark matter once the scalar field $\Phi $ oscillates around the minimum of the potential. Its fluctuations are also the same than those of standard cold dark matter but it predict a cut-off in the Mass Power Spectrum, now due to its quadratic behavior. This results are enclosed in the following relations\cite{luis,luis2}

\begin{eqnarray}
V_0 &\simeq& \frac{1.7}{9} \left( \frac{\Omega_{0CDM}}{\Omega_{0\gamma}} \right)^3 \left( \lambda^2 -4 \right)^3 \rho_{0CDM}, \nonumber \\
m^2_{\Phi} &=& \kappa_0 V_0 \lambda^2 , \label{util2} \\
k_{J} &\simeq& 1.3 \, \lambda \sqrt{\lambda^2-4} \, \frac{\Omega_{0CDM}}{\sqrt{\Omega_{0\gamma}}} H_0.  \nonumber
\end{eqnarray}

\noindent $\Omega _{0CDM}$ and $\Omega _{0\gamma }$ are the current contributions of dark matter and radiation to the critical energy density $\rho _{crit}=3H_{0}^{2}/\kappa _{0}$ with $H_{0}$ the current Hubble parameter and then $\rho_{0CDM}= \Omega_{oCDM} \rho_{crit}$. $m_{\Phi }^{2}$ is the mass of the scalar field and $k_{J}$ is the Jeans wave number at which the Mass Power Spectrum has a cut-off, a very important difference with respect to the standard cold dark matter model.

We recall that the model, at this level, has only one free parameter: $\lambda $. Nevertheless, if we expect a cut-off in the Mass Power Spectrum about the wave number $k\simeq 4.5 \, h \,Mpc^{-1}$\cite{liddle}, we can use the last of eqs. (\ref{util2}) to determine $\lambda$. It turns out that $\lambda \simeq 20.28$, then, we also obtain that $V_0\simeq \left(3.0\times 10^{-27}\,M_{Pl}\simeq 36.5\,eV\right) ^{4}$ and the corresponding ultra-light mass of the scalar field is $m_{\Phi }\simeq 9.1\times 10^{-52}\,M_{Pl}\simeq 1.1\times 10^{-23}\,eV$\cite{luis2}. In other words, the expected cut-off of the Mass Power Spectrum in the Universe fixes the last free parameter of the model. Up to this point, all parameters in potential (\ref{cosh}) are completely determined, we recover the success of standard cold dark matter model at large scales and we have in addition a cut-off in the Mass Power Spectrum, alleviating problems of the standard model at subgalactic level.

Some differences with respect to other scalar field models appear because of the non-polynomial behavior of (\ref{cosh}) in certain regimes. For instance, at cosmological scale, the exponential-like behavior provokes no problem with initial conditions either for $\rho _{\Phi }$ or the fluctuations (no fine-tuning). Also, the Jeans scale is related to the time at which the scalar field changes its behavior as radiation-like to cold dark matter-like\cite{luis2}, not to the time of radiation-matter equality as in polynomial potentials\cite{hu}. 

Moreover, because of the presence of a scalar field potential (\ref{cosh}), there must be an important self-interaction among the scalar particles. For example, even if scalar particles are lighter than neutrinos, the former can cluster due to the presence of the scalar potential. That means that this scalar field dark matter model belongs to the so called group of self-interacting dark matter models which are characterized by a $2\rightarrow 2$ scattering cross section\cite{stein,stein2}.

\section{Renormalization and $2\rightarrow 2$ scattering cross section}

The scalar potential (\ref{cosh}) can be written as a series of even powers in $\Phi $, $V=\sum_{n=1}^{\infty }V_0\lambda ^{2n}\kappa _{0}^{n}\Phi^{2n}$. Working on 4 dimensions, it is commonly believed that only $\Phi ^{4}$ and lower order theories are renormalizable. But, following the important work\cite{halpern}, if we consider that there is only one intrinsic scale $\Lambda $ in the theory, we conclude that there is a momentum cut-off and that we can have an effective $\Phi^{4}$ theory which depends upon all couplings in the theory. In addition, it was recently demonstrated\cite{bran} that scalar exponential-like potentials of the form (we have used the notation of potential (\ref{cosh}) and for example, parameter $\mu $ in eqs. 7-8 in ref.\cite{bran} is $\mu ^{-1}=\lambda \sqrt{\kappa _{0}}$))

\begin{equation}
U_\Lambda (\Phi) = M^4 \exp{\left( - \frac{\lambda^2 \kappa_0 \Lambda^2}{32
\pi^2} \right)} \exp{\left( \pm \lambda \sqrt{\kappa_0} \Phi \right)}
\label{renpot}
\end{equation}

\noindent are non-perturbative solutions of the exact renormalization group equation in the Local Potential Approximation (LPA) (we have taken a massless gaussian fixed point potential, i.e. parameter $m^{2}=0$ in eq. (5) of \cite{bran}). Here $M$ and $\lambda $ are free parameters of the potential and $\Lambda$ is the scale of renormalization. Being our potential a $\cosh $-like potential (non-polynomial), it is then a solution to the renormalization group equations in the (LPA), too. Comparing eqs. (\ref{cosh}, \ref{renpot}), we can identify $V_0=M^4 \tau$ with

\begin{equation}
\tau = \exp{\left( - \frac{\lambda^2 \kappa_0 \Lambda^2}{32 \pi^2} \right)}
\label{tau}
\end{equation}

\noindent and we see that an additional free parameter appears, the scale of renormalization $\Lambda$. From this, we can assume that potential (\ref{cosh}) is renormalizable with only one intrinsic scale: $\Lambda$. Even though the parameters in the scalar potential (\ref{cosh}) were fixed by cosmological observations, the scale $\Lambda$ remains free.

We now proceed to calculate the $2 \rightarrow 1$ scattering cross section for the scalar particles. Following the procedure shown in\cite{halpern2} for a potential with even powers of the dimensionless scalar field $\phi$, $U(\phi)=\sum^{\infty}_{n=1} u_{2n} \phi^{2n}$, the scattering amplitudes are given by

\begin{equation}
A_{2n} = \sum^{\infty}_{m=n} u_{2m} \left( \frac{I}{2} \right)^{m-n} \frac{(2m)!}{(m-n)!},  \label{series}
\end{equation}

\noindent where, in our case, the dimensionless $u_{2m}$ are

\begin{equation}
u_{2m} = \left( \frac{M}{\Lambda} \right)^4 \frac{\tau \lambda^{2m} \left( \Lambda^2 \kappa_0 \right)^m }{(2m)!},
\end{equation}

\noindent and the internal contraction $I$ reads

\begin{equation}
I \equiv \int^1_0 \frac{d^4 k}{(2\pi)^4} \frac{1}{k^2+2u_2}= \frac{1}{16\pi^2} \left( 1+ 2u_2 \ln{\frac{2u_2}{1+2u_2}} \right).  \label{int}
\end{equation}

\noindent Summing the series (\ref{series}) we find

\begin{equation}
A_{2n} = \left( \frac{M}{\Lambda} \right)^4 \lambda^{2n} \left( \Lambda^2 \kappa_0 \right)^n \tau \exp{\left( \frac{I \lambda^2 \kappa_0 \Lambda^2}{2} \right)}. \label{amp}
\end{equation}

\noindent Since the only scale that appears naturally in gravitation is the Planck mass in $\kappa_0$, we can expect that $\Lambda \sim {\mathcal O}(M_{Pl})$ and then $2 u_2 = (m^2_\Phi/ \Lambda^2) \ll 1$. In consequence, $I \simeq 1/(16 \pi^2)$. Then, the cross section for $2 \rightarrow 2$ scattering in the center-of-mass frame is

\begin{equation}
\sigma_{2 \rightarrow 2} = \frac{g^2}{16 \pi E^2} \exp{\left( \frac{\lambda^2 \kappa_0 \Lambda^2}{16 \pi^2} \right)},  \label{sigma1}
\end{equation}

\noindent where $E$ is the total energy. At this point, we can notice the non-perturbative nature of potential (\ref{cosh}). Had we taken perturbatively the scalar potential (\ref{cosh}) with a coupling $g=4! u_4 \simeq 2 \times 10^{-97}$, the self-interaction would have appeared as extremely weak. But, surprisingly, we can observe from eqs. (\ref{tau}, \ref{amp}) that the contribution of higher-order couplings points to an effective coupling $g_{eff} = g \tau^{-1}$.

\section{Planck-scale physics and galactic consequences}

An interesting dark matter model is that of a self-interacting dark matter\cite{stein,stein2}. This proposal considers that dark matter particles have an interaction characterized by a scattering cross section by mass of the particles given by \cite{stein,avila}

\begin{equation}
\frac{\sigma_{2 \rightarrow 2}}{m} = 10^{-25}-10^{-23} cm^2 \, GeV^{-1}.
\label{sigma}
\end{equation}

\noindent This self-interaction provides shallow cores of galaxies and a minimum scale of structure formation, that it must also be noticed as a cut-off in the Mass Power Spectrum\cite{avila2,liddle}. The cosh-potential (\ref{cosh}) directly provides this cut-off, but the self-interaction among scalar particles would be important for other aspects of galaxy evolution.

We can guess the value of the scale $\Lambda$ by considering the expected result (\ref{sigma}). Near the threshold $E \simeq 4m^2$ and taking the values of $\lambda$, $m_\Phi$, we find that

\begin{equation}
\Lambda \simeq (1.93 \pm 0.01) \, M_{Pl} \simeq 2.3 \times 10^{19} \, GeV.
\end{equation}

\noindent The range of $\Lambda$ is very narrow because of the exponential behavior. Observe that the scale of renormalization is of order of the Planck Mass. Recalling that $ V_0 \tau^{-1}=M^4$, we also find that $M= (6.7 \pm 1.9) \times 10^2$ TeV. All parameters are completely fixed now.

An intriguing point is the appearance of two different energy scales, one in $\Lambda$ and another one in $M$ in (\ref{renpot}), and it could be not only a coincidence that the former is of ${\mathcal O}(M_{Pl})$ and the latter is of ${\mathcal O}(100\,TeV)$. It should be noted that it is the combination of these scales which determines the observable value of parameters $V_0$ and $m_{\Phi }$, at the cosmological scale, and the value of $\sigma $ at the galactic one. This could be the first case in which the relevant scales are hidden into the effective theory at low-energies, as it was claimed that the solution (\ref{renpot}) is a solution of the renormalization group equations ``irrespectively of the high energy theory''\cite{bran}. It is also worth mentioning that the potential (\ref{cosh}) has been widely studied in non perturbative field theory in $1+1$ dimensions where it is known as the sinh-Gordon model (see \cite{saleur} and references therein).

We would like to mention that our first motivation was to investigate the hypothesis of scalar dark matter in galaxies, from a pure general relativistic point of view\cite{siddh}. The main result have been that an exponential-like potential appears naturally if one wants to fit rotation curves in galaxies, in both axially\cite{siddh} and spherically symmetric cases\cite{siddh2}. Since it is known that the formation of galaxies from dark matter involves non-lineal phenomena that can only be analyzed using numerical simulations, the cases treated in\cite{siddh,siddh2} can be considered as good approximations to the exact solution with the potential (\ref{cosh}).

But, even if the evolution of a galaxy could be described by self-interacting dark matter models\cite{stein2}, the scalar nature of dark matter can provide us with a more interesting picture: Bose condensation\cite{jeremy,tkachev}. For instance, the relaxation time for the condensate would be smaller than the age of the Universe if\cite{tkachev}

\begin{equation}
g_{eff} > 6 \times 10^{-15} \left( m_{\Phi}/eV \right)^{7/2}.
\end{equation}

\noindent We find that $\Lambda > 1.72 \, M_{Pl}$. Again, $\Lambda$ should be of order of the Planck mass. From this, it is clear that, at galactic scales, the scalar field dark matter model with the scalar potential (\ref{cosh}) must be studied both in general relativity and quantum mechanics. This is beyond the purpose of this paper and is left for future work.

Summarizing, a scalar field endowed with the scalar potential (\ref{cosh}) is a reliable model as dark matter in the Universe, not only at cosmological level, but also at galactic level because of its self-interaction. Even if we are dealing with a non-perturbative case, the scalar potential is renormalizable at a energy scale of order of the Planck mass. Also, all relevant scales are hidden into the parameters of the effective theory, but they might be very important when performing quantum calculations.

\ack{We would like to thank F. Siddhartha Guzm\'an and Dar\'{\i}o N\'u\~{n}ez for many helpful discussions. L.A.U. thanks Rodrigo Pelayo and Juan Carlos Arteaga Vel\'{a}zquez and T.M. thanks Gabriel L\'opez Castro for very helpful comments. This work was partly supported by CONACyT, M\'{e}xico 119259 (L.A.U.).}


\end{document}